\documentclass[prb,twocolumn,showpacs,preprintnumbers,amsmath,amssymb]{revtex4}
\usepackage{graphicx}% Include figure files
\usepackage{dcolumn}% Align table columns on decimal point
\usepackage{bm}% bold math
\usepackage{latexsym,epsfig}

\begin{document}
\preprint{version final}

\title{Phases and Transitions in the Spin-1 Bose-Hubbard Model:
Systematics of a Mean-field Theory}
\author{Ramesh V. Pai}
\email{rvpai@unigoa.ac.in}
\affiliation{
Department of Physics, Goa University, Taleigao Plateau,
Goa 403 206, India.
}
\author{K. Sheshadri}
\email{kshesh@gmail.com} \affiliation{ 686, BEL Layout, 3$^{rd}$
Block, Vidyaranyapura, Bangalore 560 097, India. }
\author{Rahul Pandit}
\email{rahul@physics.iisc.ernet.in}
\homepage{http://www.physics.iisc.ernet.in/~rahul}
\altaffiliation [also at ]{Jawaharlal Nehru Centre for Advanced
Scientific Research, Jakkur, Bangalore 560 064, India.
}
\affiliation{
Centre for Condensed Matter Theory, Department of Physics,
  Indian Institute of Sciences,
  Bangalore 560 012, India.
}
\date{\today}
\begin{abstract}
We generalize the mean-field theory for the spinless Bose-Hubbard
model to account for the different types of superfluid phases that
can arise in the spin-1 case. In particular, our mean-field theory
can distinguish polar and ferromagnetic superfluids, Mott insulators
which arise at integer fillings at zero temperature, and normal Bose
liquids into which the Mott insulators evolve at finite
temperatures. We find, in contrast to the spinless case, that
several of the superfluid-Mott insulator transitions are first-order
at finite temperatures. Our systematic study yields rich phase
diagrams that include, first-order and second-order transitions, and
a variety of tricritical points. We discuss the possibility of
realizing such phase diagrams in experimental systems.

\end{abstract}

\maketitle

\section{Introduction}
Experimental investigations of ultracold atoms in optical lattices
have opened up a new realm in the study of quantum phase transitions
(QPT)~\cite{review,sachdev}. The  superfluid (SF) to Mott-insulator
(MI) transition has been observed in spin-polarized  $^{87}$Rb atoms
trapped in a three-dimensional, optical-lattice
potential~\cite{greiner}, by changing the strength of the onsite
potential, as predicted theoretically by studies of the spinless
Bose-Hubbard model~\cite{jaksch,sheshadri}. Furthermore, technical
advances in the trapping of atoms by purely optical
means~\cite{stamper} have enhanced the interest in the study of
quantum magnetism in confined dilute atomic gases. Alkali atoms with
nuclear spin $I=3/2$, such as $^{23}Na$, $^{39}K$, and $^{87}Rb$,
have hyperfine spin $F=1$. In conventional {\it magnetic traps},
these spins are frozen, so the atoms can be treated as spinless
bosons; by contrast, in {\it purely optical traps}, these spins are
free, so the Bose condensates, which form at low temperatures, can
have a spinor nature~\cite{ho1,muker} and the SF-MI transition can
be modified~\cite{tsuchiya,graham,rvpiacs}. In the spinless case,
the SF-MI transitions are controlled by the interaction $U_0$
between bosons at the same site. As $U_0$ increases beyond a
critical value, the SF phase undergoes a continuous transition to an
MI phase in which the number of bosons at every site is an integer.
This transition is reflected in the development of a gap at the
transition. When the spin is nonzero such a gap also develops at
SF-MI transitions, but the properties of the phases and the natures
of these transitions are modified by the spin degrees of freedom.

Theoretical work on this problem has dealt primarily with the
properties of spinor condensates by using a continuum, effective,
low-energy Hamiltonian. Such a Hamiltonian suffices if one is
interested in the natures of the superfluid phases, which can be
polar or ferromagnetic, and in their excitations, which include
vector or quadrupolar spin waves and topological
defects\cite{ho1,muker}. However, if we want to study the SF-MI
transitions we must use a lattice model such as the spin-one
Bose-Hubbard model. Some groups~\cite{tsuchiya,graham,rvpiacs} have
initiated such an investigation by obtaining the zero-temperature
phase diagram of this model in a mean-field approximation. The
topology of this phase diagram for the spin-1 Bose-Hubbard model is
similar to that of its spinless counterpart; but,
in the spin-1 model,
the superfluid phases can be either polar or ferromagnetic depending
on whether the spin-dependent interaction favors or disfavors the
formation of singlets. In the former case the SF-MI phase transition
is {\it continuous}, if the density of bosons per site $\rho$ is
an odd number, but {\it first-order}, if $\rho$ is an even number.

We have two main goals in this paper: The first is to give a global
view of the zero-temperature, mean-field-theory phase diagram of the
spin-1 Bose Hubbard model emphasizing issues of first-order
coexistence that have not been highlighted so far. The second is to
generalize this mean-field theory to finite temperatures $T > 0$ and
thus obtain the finite-temperature, mean-field-theory phase diagram
for this model.

The spin-1 Bose-Hubbard model is defined by the Hamiltonian
\begin{eqnarray}
\label{eq:bh}
{\cal{H}}=&&
-t \sum_{<i,j>,\sigma}(a^{\dagger}_{i,\sigma}a_{j,\sigma}+
h.c) + \frac{U_0}{2} \sum_i {\hat{n}_i}({\hat{n}_i}-1) \nonumber \\
&&+\frac{U_2}{2} \sum_i (\vec{F}^2_i - 2\hat{n_i})-\sum_{i}\mu_i
\hat{n_i},
\end{eqnarray}
where the first is the kinetic energy associated with the hopping of
bosons between nearest-neighbor pairs of sites $<i,j>$ with
amplitude $t$, $a_{i,\sigma}^\dagger$ ($a_{i,\sigma}$) is the boson
creation (annihilation) operator at site $i$ with spin component
$\sigma$ (which can assume the values $1,0,-1$),
$\hat{n_{i\sigma}}\equiv a^{\dagger}_{i,\sigma}a_{i,\sigma}$;
$\hat{n_i}\equiv\sum_\sigma\hat{n_{i,\sigma}}$ and
$\vec{F}_i=\sum_{\sigma,\sigma '} a^{\dagger}_{i,\sigma}
\vec{F}_{\sigma ,\sigma '} a_{i,\sigma '}$ are, respectively, the
total boson number and spin on site $i$, and $ \vec{F}_{\sigma
,\sigma '}$ are the standard spin-one matrices; $U_0$ is the onsite
Hubbard repulsion and $U_2$ the energy for nonzero spin
configurations on a site. The origin of such a spin-dependent term
lies in the difference between the scattering lengths $a_0$ and
$a_2$, for $S=0$ and $S=2$ channels~\cite{law}, respectively; in
terms of these lengths $U_0=4\pi\hbar^2(a_0+2 a_2)/3M$ and
$U_2=4\pi\hbar^2(a_2- a_0)/3M$, where $M$ is the mass of the
atom~\cite{ho1}. For $^{23}Na$, $a_2=54.7 a_B$ and $a_0=49.4 a_B$,
where $a_B$ is the Bohr radius, so $U_2 > 0$, whereas for $^{87}Rb$,
$a_2=(107\pm 4) a_B$ and $a_0=(110\pm 4) a_B$, so $U_2 $ can be
negative. The parabolic trapping potential with strength $V_T$ is
represented by the site-dependent chemical potential
$\mu_i=\mu-V_T|R_i|^2$, where $R_i$ is the distance of site $i$ from
the center of the trap and $\mu$ is a uniform chemical potential
that controls the mean density of the bosons. In this study we
neglect the trap potential (i.e., we set $V_T=0$) and focus on the
effects of the spin degrees of freedom.

The zero-temperature phase diagram of model (\ref{eq:bh}) has been
obtained in the mean-field approximation by some
groups~\cite{tsuchiya,graham,rvpiacs}. We have extended these
studies significantly. Before presenting the details of our
work, we give a qualitative overview of our new results.

Consider first the case $U_2=0$: We might expect
the spin-1 and spinless Bose-Hubbard models to have same
phase diagram in this case since the ground-state energy does
not depend on the spin. This is superficially true at $T = 0$
in so far as the SF-MI phase boundaries for both these
models overlap. However, as we will show, the SF phase is
highly degenerate in the spin-1 case; and for $T > 0$ the
SF-MI transition becomes first-order and, eventually,
continuous again. Thus the finite-temperature phase
diagram has a rich topology with first-order boundaries
evolving into continuous ones at tricritical points.

If $U_2\neq 0$ the onsite interaction between the bosons becomes
spin dependent. It turns out that we must distinguish between the
cases $U_2 < 0$ and $U_2 > 0$. The former yield a phase diagram that
is very similar to the one for $U_2 = 0$; the major qualitative
difference arises in the nature of the SF phase that is now a
ferromagnetic superfluid.

There are many differences between the phase diagrams of
the spin-1 model with $U_2 = 0$ and $U_2 > 0$. If
$U_2 > 0$ the SF phase is a polar superfluid. Furthermore,
even at $T = 0$, the SF-MI transitions are different for odd
and even densities. For odd densities, the $T = 0$ SF-MI transition
is continuous as for $U_2=0$; however, for even
densities, this SF-MI transition turns out to be first-order
because of the formation of singlets that also stabilize
the MI phase considerably.

At finite temperatures the MI phases evolve without a singularity
into a normal Bose liquid (NBL). These are really not distinct
phases but, as we will show, the compressibility $\kappa$ can be
used effectively to delineate the crossover between MI and NBL
regions.

To present our results in detail we must introduce our
mean-field theory. We do this in Sec. II.  Our results are given
in Sec. III. We end with a discussion in Sec. IV.

\section{Mean-field Theory}

Mean-field theory has been very successful in obtaining the phase
diagram for the spinless Bose-Hubbard model. There are three
formulations of this mean-field theory: one uses a model with
infinite-range interactions, another a Gutzwiller-type wave
function, and a third\cite{sheshadri}, which we follow, a decoupling
approximation. The unique feature of this decoupling scheme is that,
unlike conventional mean-field theories, it does not decouple the
interaction term to obtain an effective, \emph{one-particle }
problem but, instead, decouples the hopping term to obtain an
effective, \emph{one-site} problem. This one-site problem is then
solved self-consistently. We generalize this decoupling procedure to
the spin-1 case as follows\cite{sheshadri}: In the identity
$a^{\dagger}_{i,\sigma}a_{j,\sigma} = (a^{\dagger}_{i,\sigma} -
\langle a^{\dagger}_{i,\sigma}\rangle) (a_{j,\sigma} - \langle
a_{j,\sigma}\rangle) + \langle a^{\dagger}_{i,\sigma}\rangle
a_{j,\sigma} +a^{\dagger}_{i,\sigma} \langle a_{j,\sigma}\rangle
-\langle a^{\dagger}_{i,\sigma}\rangle \langle a_{j,\sigma}\rangle$,
where $\langle {\cal{O}} \rangle$ denotes the equilibrium value of
an operator ${\cal{O}}$, we neglect the first term that is quadratic
in deviations from the equilibrium value. Thus
\begin{eqnarray}
\label{eq:hop}
a^{\dagger}_{i,\sigma}a_{j,\sigma} &\simeq& \langle
a^{\dagger}_{i,\sigma}\rangle a_{j,\sigma}
+a^{\dagger}_{i,\sigma}\langle a_{j,\sigma}\rangle \nonumber \\
&&-\langle a^{\dagger}_{i,\sigma}\rangle \langle a_{j,\sigma}\rangle
;
\end{eqnarray}
since we expect superfluid phases, it is natural to introduce the
superfluid order parameters
\begin{equation}
\label{eq:opar}
\psi_{\sigma}\equiv \langle a^{\dagger}_{i,\sigma}\rangle
\equiv \langle a_{i,\sigma}\rangle ,
\end{equation}
for $\sigma = 1, 0, -1$. We consider equilibrium states
with uniform phases, so we choose these order parameters to be real.
Given the decoupling approximation (\ref{eq:hop}) the
Hamiltonian~(\ref{eq:bh}) can be written as a sum over single-site,
mean-field Hamiltonians:
\begin{equation}
\label{eq:bh_site}
 {\cal{H}}=\sum_{i}  {\cal{H}}_i^{MF},
\end{equation}
where
\begin{eqnarray}
\label{eq:mfh}
 {\cal{H}}_i^{MF}&=& \frac{U_0}{2} {\hat{n}_i}({\hat{n}_i}-1)
+\frac{U_2}{2} (\vec{F}^2_i - 2\hat{n_i})-\mu \hat{n_i}\nonumber \\
&& -\psi_{\sigma}(a^{\dagger}_{i,\sigma}+a_{i,\sigma}) +
 \sum_{\sigma}|\psi_\sigma|^2.
\end{eqnarray}
We set the energy scale by choosing $zt=1$, where $z$ is the number
of nearest neighbors. At least one of the order parameters
$\psi_{\sigma}$ is nonzero in a superfluid phase. In order to
calculate $\psi_\sigma$ in our mean-field theory, we first obtain
the matrix elements of the mean-field Hamiltonian ${\cal{H}}_i^{MF}$
in the onsite, occupation-number basis $\{|n_{-1},n_0,n_1>\}$
truncated at a finite value $n_{\mbox{max}}$ of the total number of
bosons per site $n=\sum_{\sigma} n_\sigma$. In most of our
studies we use $n_{max}=4$ for which the mean-field Hamiltonian is a
$36 \times 36$ matrix \cite{foot1}.  We diagonalize this
matrix to obtain its eigenvalues ${\cal E}_\alpha$ and eigenvectors
$\mid \varphi_\alpha \rangle$:
\begin{equation}
\label{eq:spectrum}
 {\cal{H}}_i^{MF} |\varphi_\alpha \rangle   =
{\cal E}_\alpha  |\varphi_\alpha \rangle;
\end{equation}
we suppress the site index $i$ on these eigenvalues and eigenvectors
since all the phases we consider are spatially uniform.

\begin{figure*}[htbp]
\centering \epsfig{file=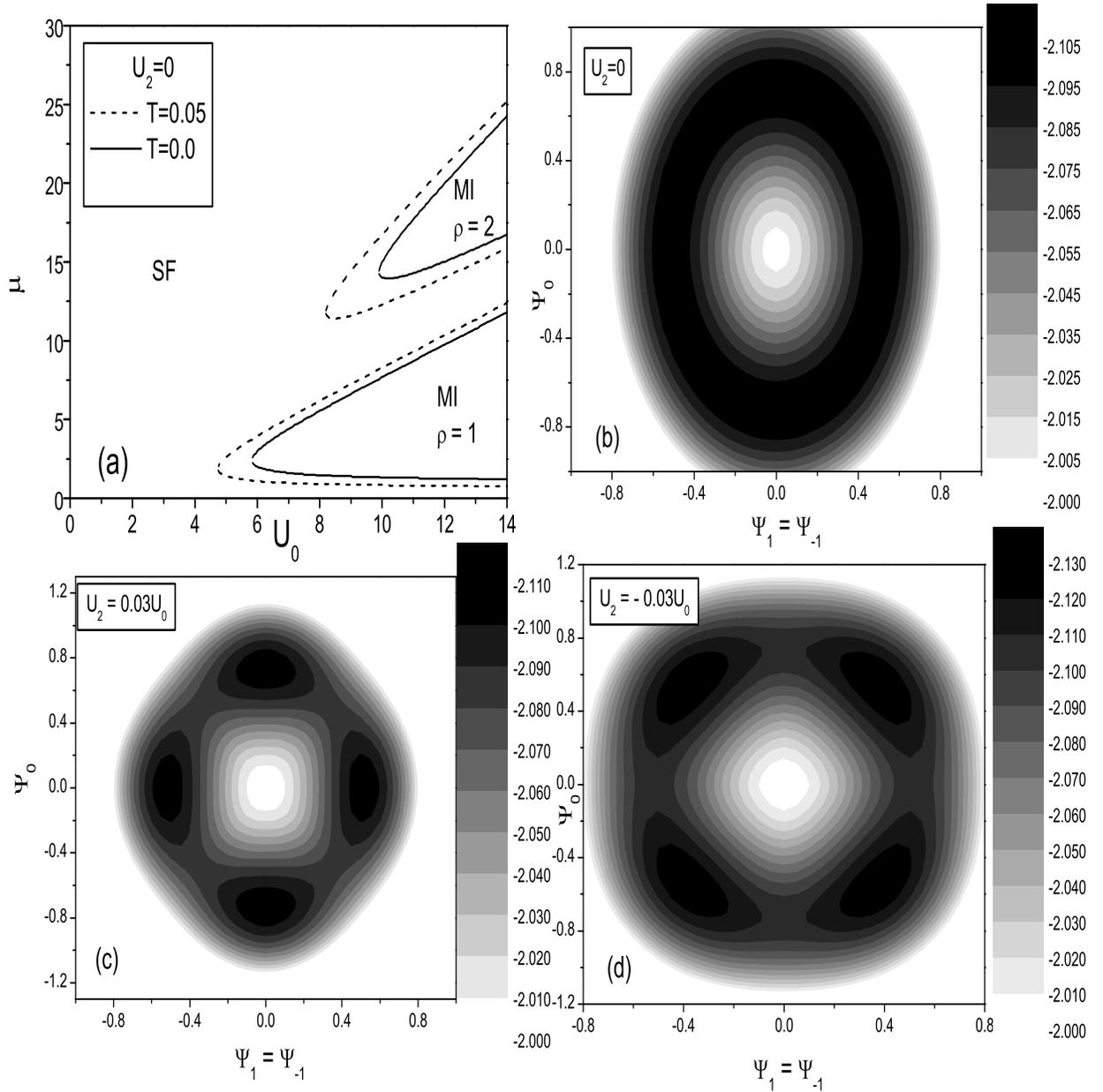,width=20cm,height=20cm}
\caption{(a) Phase diagrams in the ($\mu-U_0$) plane for $U_2=0$:
Solid lines indicate the $T=0$ continuous phase boundaries between
SF and MI phases; these phase boundaries evolve into first-order
boundaries at finite (but low) temperatures as shown by the
representative dashed lines for $T=0.05$; at higher temperatures
these first-order boundaries become continuous again at lines of
tricritical points. Pseudo-grayscale plots at $T=0$ of the
variational ground-state energy ${\cal E}_0$ for (b) $U_2=0$, (c)
$U_2/U_0=0.03$, and (d) $U_2/U_0=-0.03$, respectively; the four
degenerate minima in (c) and (d) show that the SF phase is polar in
the former and ferromagnetic in the latter; in case (b), i.e., $U_2
= 0$, the SF phase is infinitely degenerate (see text).}
 \label{fig:lobes}
\end{figure*}

We now obtain the variational the free energy
\begin{equation}
\label{eq:free}
{\cal F}(\mu, U_0, U_2, T; \psi_\sigma) =
-T \ln Z(\mu, U_0, U_2, T; \psi_\sigma),
\end{equation}
where $Z(\mu, U_0, U_2, T; \psi_\sigma)$ is the partition function
\begin{equation}
\label{eq:partfunc}
Z(\mu, U_0, U_2, T; \psi_\sigma) =
\sum_{\alpha}{e^{-{\cal E}_\alpha/T}},
\end{equation}
with the Boltzmann constant $k_B$ chosen to be $1$.
The variational free energy ${\cal F}$ must be minimized
with respect to the order parameters $\psi_\sigma$, i.e., we must
solve the equations $\partial {\cal F}/\partial \psi_\sigma = 0$ for
$\sigma = 1, 0, -1$. These equations can be recast as
self-consistency conditions for $\psi_\sigma$; solutions of these
self-consistency conditions correspond to extrema of ${\cal F}$. In
case there is more than one solution, we must pick the one that
yields the global minimum of ${\cal F}$. [At a first-order
phase boundary ${\cal F}$ has two, equally deep, global minima.]
The values of $\psi_\sigma$ and ${\cal F}$ at the global minimum
yield the equilibrium order parameters $\psi_\sigma^{eq}$ and free
energy ${\cal F}^{eq}$. In our mean-field theory, the superfluid
density is
\begin{equation}
\label{eq:rho_s}
\rho_s=\sum_\sigma \mid \psi_\sigma^{eq} \mid^2.
\end{equation}
The magnetic properties of the superfluid phases of this model are
obtained from ~\cite{ho1}
\begin{equation}
\langle\vec{F}\rangle=\frac{\sum_{\sigma,\sigma '}
\psi_{\sigma}^{eq} \vec{F}_{\sigma ,\sigma '} \psi_{\sigma '}^{eq}}
{\sum_{\sigma}|\psi_\sigma^{eq}|^2};
\end{equation}
the explicit forms of the spin-1 matrices now yield
\begin{eqnarray}
\label{eq:s_ave}
\langle\vec{F}\rangle &=& \sqrt{2}\frac{( \psi_1 \psi_0 + \psi_{-1}
\psi_0)} {\sum_{\sigma}|\psi_\sigma|^2} \hat{x} +
\frac{(\psi_1^2-\psi_{-1}^2)}{\sum_{\sigma}|\psi_\sigma|^2} \hat{z},
\nonumber \\
\langle\vec{F}\rangle^2 &=& 2 \frac{( \psi_1 \psi_0 + \psi_{-1}
\psi_0)^2} {(\sum_{\sigma}|\psi_\sigma|^2)^2}
+\frac{(\psi_1^2-\psi_{-1}^2)^2}{(\sum_{\sigma}|\psi_\sigma|^2)^2},
\end{eqnarray}
where $\hat{x}$ and $\hat{z}$ are unit vectors in spin space and we
suppress the superscript $eq$ for notational convenience; all
$\psi_\sigma$ used here and henceforth are actually
$\psi_\sigma^{eq}$. Superfluid states with $\langle\vec{F}\rangle=0$
and $\langle\vec{F}\rangle^2=1$ are referred to as polar and
ferromagnetic, respectively. The polar state~\cite{muker} has an
order-parameter manifold $(U(1) \times S^2)/{\mathbb{Z}}_2$, where
$U(1)$ denotes the phase angle $\theta$, $S^2$ refers to the
directions $\hat{\bf{n}}$ on the surface of a unit sphere (on which
orientations are specified by the angles $(\alpha, \beta)$ of the
spin quantization axis), and ${\mathbb{Z}}_2$ arises because of the
symmetry of this state under the simultaneous transformations
$\theta\to\theta+\pi$ and $\hat{\bf{n}} \to - \hat{\bf{n}}$. Thus
the superfluid order parameters can be written as
\begin{equation}
\label{eq:sym_plus}
\left(
\begin{tabular}{c}
$\psi_1$ \\
$\psi_0$ \\
$\psi_{-1}$
\end{tabular}
\right)
=
\sqrt{\rho_s}
 e^{\imath \theta}\left(
\begin{tabular}{c}
$-\frac{1}{\sqrt{2}}e^{-\imath \alpha} \sin\beta$ \\
$ \cos\beta$ \\
$\frac{1}{\sqrt{2}}e^{\imath \alpha} \sin\beta$. \\
\end{tabular}
\right).
\end{equation}
Similarly, since the ferromagnetic superfluid state has an
order-parameter manifold~\cite{muker} with the symmetry group
$SO(3)$,
\begin{equation}
\label{eq:sym_minus}
\left(
\begin{tabular}{c}
$\psi_1$ \\
$\psi_0$ \\
$\psi_{-1}$
\end{tabular}
\right)
=
\sqrt{\rho_s}
 e^{\imath (\theta-\tau)}\left(
\begin{tabular}{c}
$e^{-\imath \alpha} \cos^2\frac{\beta}{2}$ \\
$ \sqrt{2}\cos\frac{\beta}{2}\sin\frac{\beta}{2}$ \\
$e^{\imath \alpha} \sin^2\frac{\beta}{2}$. \\
\end{tabular}
\right),
\end{equation}
where $\alpha, \beta$, and $\tau$ are Euler angles.

\begin{figure*}[htbp]
  \centering
  \epsfig{file=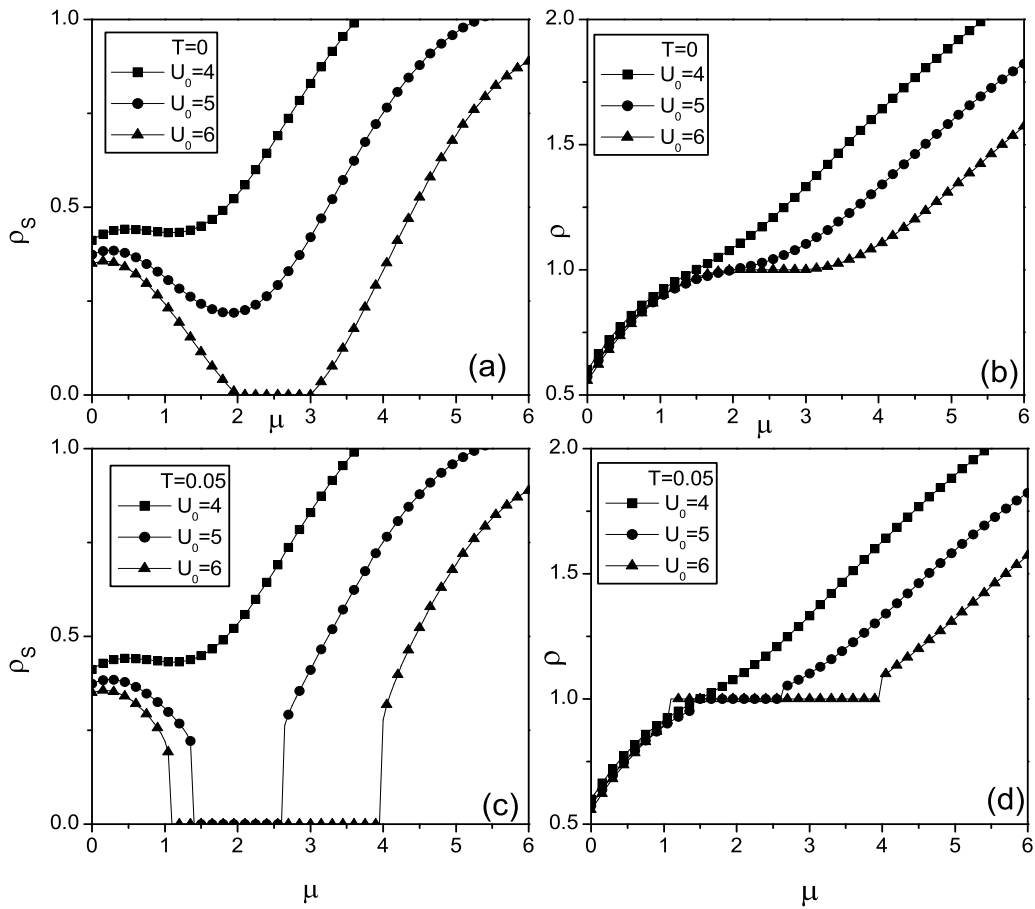,width=16cm,height=14cm}
  \caption{Representative plots of (a) $\rho_s$ and (b) $\rho$
versus $\mu$ for $U_0=4,5,6$, $U_2=0$ at $T=0$. Similar plots at
$T=0.05$ are given in (c) and (d). In the MI phases $\rho_s=0$ and
$\rho$ is an integer [$=1$ in (b) for $U_0 = 6$] for $T=0$; for $0
\lesssim T$ the MI phase evolves without a singularity (see text)
into the normal Bose liquid (NBL) in which $\rho$ is exponentially
close to an integer [$=1$ in (d) for $U_0 = 5, 6$]. As the
temperature increases from zero, the MI phases grow at the expense
of SF phase and the SF-MI transition becomes first order [see
Fig.(\ref{fig:lobes}].} \label{fig:rhos}
\end{figure*}

We consider only spatially uniform superfluids in equilibrium, so
it suffices to use real order parameters. Thus, for the
polar superfluid we have the following possibilities:
(i) $\psi_1 = \psi_{-1} > 0$ and $\psi_0 = 0$ with
$\theta = \alpha = \beta = \pi/2$ or $\theta = -\alpha = \beta =
\pi/2$; and (ii) $\psi_1 = \psi_{-1}= 0$ and $\psi_0 > 0$ with
$\beta = 0$ or $\pi$, $\theta = 0$ or $\pi$, and  $0 \leq \alpha
\leq 2 \pi$. Similarly for the ferromagnetic superfluid
$\psi_1 = \psi_{-1}$, $\beta = \pi/2$, $\alpha = 0$, $0 \leq
\theta = \tau \leq 2 \pi$, and $\psi_0 = \sqrt{2} \psi_1$.

The equilibrium density $\rho$ and compressibility $\kappa$ can be
obtained from
\begin{equation}
\label{eq:filling}
\rho = -\frac{\partial {\cal F}^{eq}}{\partial \mu}=\frac{1}{Z}
\sum_{\alpha}{e^{-E_\alpha/T}
\langle \phi_\alpha \mid \hat{n} \mid \phi_\alpha \rangle},
\end{equation}
where $E_\alpha$ and $\mid \phi_\alpha \rangle$ are
${\cal E}_\alpha$ and $\mid \varphi_\alpha \rangle$ at the global
minimum of ${\cal F}$ and
\begin{equation}
\label{eq:compress}
\kappa = \frac{\partial \rho}{\partial \mu}.
\end{equation}

The three quantities $\rho_s,\; \langle \vec{F}\rangle^2,$ and
$\kappa$ together determine the thermodynamic phase of model
(1) for any point in the parameter space $\{\mu, U_0, U_2, T \}$ as
given in the Table~\ref{tab:phases}. Strictly speaking there is no
distinction between the Mott insulator (MI) and the normal Bose
liquid (NBL); the former exists at $T = 0$ and has $\kappa = 0$; it
evolves without any singularity into the NBL at $T > 0$;
at low $T$, the compressibility $\kappa$ is exponentially small in
the NBL so one can think of it as an MI phase; at high $T$, where
$\kappa$ is substantially different from $0$, it is best to think
of this phase as a normal Bose liquid. It is convenient, therefore,
to define a crossover boundary above which $\kappa$ is substantial;
we use the criterion $\kappa = \kappa_X = 0.02$ to obtain the
MI-NBL crossover boundary shown in some of our phase diagrams. We
must remember of course that this is not a strict phase
boundary and it depends on the value we choose for the crossover
compressibility $\kappa_X$.

\begin{table}
\begin{tabular}{|c|c|c|c|}
\hline
 &          &                             &          \\
Phases & $\rho_s$ &  $\langle \vec{F}\rangle^2$ & $\kappa$ \\
 &          &                             &          \\
\hline
 &          &                             &          \\
Polar Superfluid (PSF) & $>0$ & 0 & $>0$\\
&          &                             &          \\
Ferro Sperfluid (FSF) & $>0$ & 1 & $>0$\\
&          &                             &          \\
Mott Insulator  (MI) & 0    & --- & 0\\
&          &                             &          \\
Normal Bose Liquid (NBL) & 0    & --- & $>0$\\
\hline
\end{tabular}
\caption{The superfluid density $\rho_s$, $\langle \vec{F}
\rangle^2$, and the compressibility $\kappa$ in the different
phases of the spin-1 Bose-Hubbard model. The MI and NBL are really
the same phase (see text).}
\label{tab:phases}
\end{table}

\begin{figure}[htbp]
  \centering
  \epsfig{file=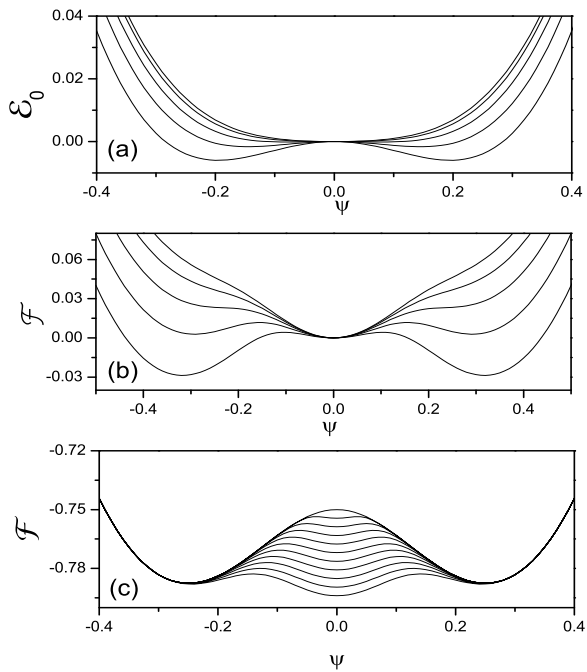,width=9cm,height=10cm}
  \caption{ Plots of the variational free energy ${\cal F}$
(ground-state energy ${\cal E}_0$ for $T=0$) as a function of $\psi$
for different values of $\mu$ in the vicinity of the SF-MI
transition for $U_2 =0$ and with $\psi_{-1}=\psi_0=\psi_1\equiv
\psi$ (see text). The minima at $\psi=0$ and $\psi\neq 0$
correspond, respectively, to MI and SF phases. (a) The two minima at
$\psi\neq 0$ merge into one minimum at $\psi=0$ to yield the
mean-field continuous SF-MI transition at $T=0$ as we increase $\mu$
from $1.5$ (bottom curve) to $2.3$ (top curve) in steps of $0.2$.
Similar plots of ${\cal F}$ are given in (b) and (c).  For
$0\lesssim T$, ${\cal F}$ develops three degenerate minima at the
SF-MI boundary, indicating clearly the coexistence of SF and MI
phases at a first-order boundary. This boundary can be crossed
either (b) by changing $\mu$ (from $0.7$ to $1.5$ in steps of $0.2$)
at fixed $T=0.05$ or (c) by changing $T$ ($0$ to $0.1$ in steps of
$0.01$) at fixed $\mu=2$. } \label{fig:free}
\end{figure}

\section{Results}

We are now in a position to present the results of our mean-field
theory. It is necessary to distinguish between three qualitatively
different regimes: (1) $U_2/U_0 = 0$; (2) $U_2/U_0 > 0$ (we use
$U_2/U_0 = 0.03$ since this is appropriate for $^{23}Na$); and (3)
$U_2/U_0 < 0$, as in $^{87}Rb$ (for specificity we use $U_2/U_0 =
- 0.03$).

\begin{figure*}[htbp]
  \centering
  \epsfig{file=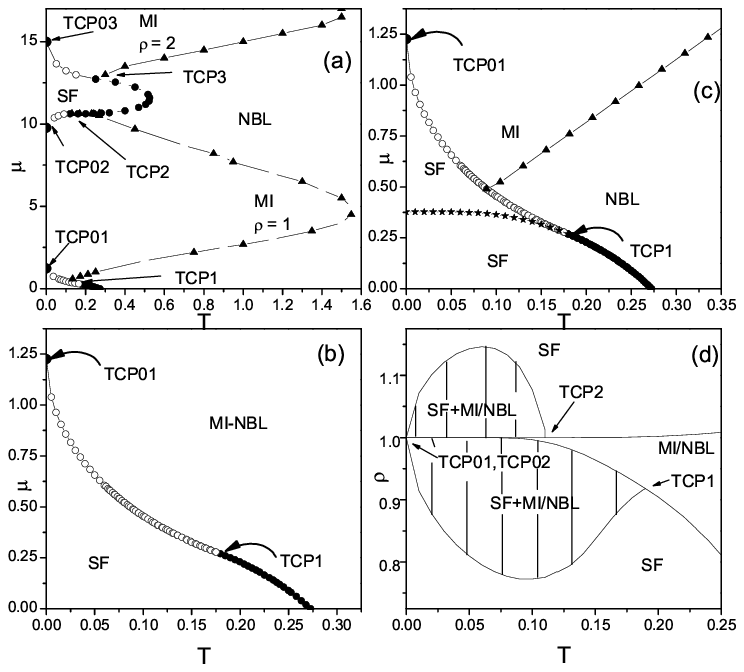,width=16cm,height=14cm}
  \caption{(a) Mean-field phase diagram in the $\mu-T$ plane for
$U_0=12$ and $U_2=0$. Lines with open (filled) circles represent
first-order (continuous) SF-MI/NBL phase boundaries. First-order
and continuous boundaries meet at tricritical points(TCP). The
$T=0$ ($T>0$) tricritical points are labeled TCP01, TCP02, etc.
(TCP1, TCP2, etc. ). The line with triangles represents the
crossover boundary between MI and NBL regions of the MI/NBL phase.
The lower left corner of the phase diagram in (a) is enlarged in
(b) and (c). The only difference between (b) and (c)
is that the latter shows the MI-NBL crossover boundary (line with
triangles) and a line with stars, the locus of points in the SF
phase at which the variational free energy ${\cal F}$ goes from a
curve with three minima to one with two minima; this line meets the
SF-MI boundary at TCP1. (d) The density-temperature $(\rho-T)$
version of part of the $\mu-T$ phase diagram of (a)
(without the MI-NBL crossover line); tie lines are used to hatch the
two-phase regions in which SF and MI/NBL phases coexist (see text).}
\label{fig:phase12}
\end{figure*}

We first consider $U_2/U_0 = 0$, which can be achieved when the
scattering lengths are equal, i.e., $a_0=a_2$. In this case the
onsite interaction between bosons is spin independent. This leads to
an infinitely degenerate superfluid state: Specifically, for a given
value of the superfluid density $\rho_s$, the three order parameters
$\psi_{\sigma}, \sigma = -1, 0, 1$, can have any magnitudes that
satisfy Eq. (\ref{eq:rho_s}); e.g., if we make the specific choice
$\psi_{-1}=\psi_1$, then the pseudo-grayscale plot of
Fig.~\ref{fig:lobes} (b) shows that the minima of the variational
mean-field energy at $T=0$ lie on the ellipse $2\psi_1^2 + \psi_0^2
= \rho_s$. This degeneracy makes the superfluid phase of the spin-1
Bose-Hubbard model different from its spin-0 counterpart {\it even
if $U_2 = 0$}; and it implies that an infinite number of SF phases
{\it coexist} at $U_2=0$. However, the zero-temperature phase
diagram of Fig.~\ref{fig:lobes} (a) is the same as that of the
spinless Bose-Hubbard model~\cite{sheshadri}; and, in particular,
lobes of the MI phase are separated from the SF phase by the SF-MI
boundaries that are all continuous at $T=0$; the density $\rho$ is
fixed at integral values in each MI lobe.

Striking differences between the spin-1 and spinless cases appear at
finite temperatures. We demonstrate this in Figs.~\ref{fig:lobes}(a)
and \ref{fig:rhos}; the former compares phase diagrams at $T=0$ and
$T=0.05$ and the latter presents plots, both at $T=0$ and $T=0.05$,
of the superfluid density $\rho_s$ and the density $\rho$ as
functions of the chemical potential $\mu$ for three different values
of the onsite interaction $U_0$ ($=4,5,6$). [In both spinless and
spin-1 cases, if $U_2 = 0$, the tip of the first
lobe~\cite{sheshadri,rvpiacs} lies at $U_{0c}(\rho=1) \simeq 5.8$
for $T=0$.] Figure~\ref{fig:lobes} (a) shows that $U_{0c}(\rho)$
decreases as the $T$ increases, i.e., the MI lobes grow at the
expense of the SF phase. Figure~\ref{fig:rhos} shows that $\rho_s$
goes to zero continuously at the SF-MI transition, if $T=0$, but
with a jump if $T=0.05$. Thus the SF-MI transition becomes a
first-order transition at finite $T$ and the zero-temperature SF-MI
boundaries [Fig. \ref{fig:lobes} (a)]are really lines of tricritical
points; as the temperature is increased further, the first-order
transition again becomes continuous at another tricritical point.

The first-order, SF-MI coexistence boundary is associated with the
three-degenerate-minima structure in the variational-free-energy
plots shown in Fig.~\ref{fig:free}. To obtain these plots we use
$\psi_{-1}=\psi_0=\psi_1\equiv\psi$ that is one of the admissible
solutions in the infinitely degenerate SF phase for the case
$U_2=0$; the infinite degeneracy of this phase, illustrated for
$T=0$ in Fig. \ref{fig:lobes} (b), persists in our mean-field theory
even if $T > 0$. Figure~\ref{fig:free} shows plots of the
variational free energy ${\cal F}$ (ground-state energy ${\cal E}_0$
for $T=0$) as a function of $\psi$ for different values of $\mu$ in
the vicinity of the SF-MI transition; the minima at $\psi=0$ and
$\psi\neq 0$ correspond, respectively, to MI and SF phases.  At
$T=0$ the SF-MI transition is continuous: this is reflected in the
plots of ${\cal E}_0$ in Fig. \ref{fig:free} (a), where, as we go
from the SF to the MI phase by changing $\mu$, two global minima
with $\psi\neq 0$ merge to yield one minimum at $\psi=0$; precisely
at the mean-field critical point we have a {\it quartic} minimum.
For $0\lesssim T$, ${\cal F}$ develops three degenerate minima at
the SF-MI boundary, indicating clearly the coexistence of SF and MI
phases at a  first-order boundary. This boundary can be crossed
either by changing $\mu$ at fixed $T$ [Fig. \ref{fig:free} (b)] or
by changing $T$ at fixed $\mu$ [Fig. \ref{fig:free} (c)]. At
sufficiently high temperatures this three-minima structure of ${\cal
F}$ goes away at a tricritical point at which the three minima
coalesce to yield a {\it sixth-order} minimum. Beyond this
tricritical point the SF-MI transition is continuous (second-order).

Calculations such as those summarized in the plots of Fig.
\ref{fig:free} help us to obtain the phase diagrams shown in Figs.
\ref{fig:phase12} (a) - (d) for $U_0 = 12$ and $U_2 =0$. Let us
begin with the $\mu-T$ phase diagram shown in Fig. \ref{fig:phase12}
(a). The MI phases [lobes in Fig. \ref{fig:lobes} (a)] at $T=0$
evolve without any singularity into the normal Bose liquid (NBL) for
$T > 0$. As we have emphasized earlier,  MI and NBL phases are not
distinct, but it is useful to think of a smooth crossover from one
to the other; we define these crossover boundaries as the loci of
points at which the compressibility $\kappa = \kappa_x=0.02$. The
MI-NBL crossover boundaries (lines with filled triangles) are also
shown in Figs. \ref{fig:phase12} (a) and (c). Islands of the SF
phase appear in the $\mu-T$ phase diagram; the first two of these
are shown in Fig. \ref{fig:phase12} (a), where one is marked SF and
the other, near the origin, is shown magnified in Figs.
\ref{fig:phase12} (b) and (c). The only difference between Figs.
\ref{fig:phase12} (b) and (c) is that the latter shows the MI-NBL
crossover boundary (line with triangles) and a line with stars, the
locus of points in the SF phase at which the variational free energy
${\cal F}$ goes from a curve with three minima to one with two
minima. This line meets the SF-MI boundary at a tricritical point
labeled TCP1. Higher islands of the SF phase show analogous
tricritical points labeled TCP2, TCP3, etc.; the SF-MI phase
boundaries meet the $T=0$ axis at the zero-temperature tricritical
points TCP01, TCP02, TCP03, etc. [Fig. \ref{fig:phase12} (a)].
Figure \ref{fig:phase12} (d) shows the density-temperature
$(\rho-T)$ version of part of the $\mu-T$ phase diagram of Fig.
\ref{fig:phase12} (a) (without the MI-NBL crossover line); the
first-order parts of the SF-MI boundaries now appear as regions of
two-phase coexistence that are hatched with tie lines; the two-phase
regions corresponding to the two lowermost SF-MI boundaries in Fig.
\ref{fig:phase12} (a) are depicted; they end at the tricritical
points TCP1 and TCP2 out of which emerge the continuous
(second-order) SF-MI phase boundaries. We use the label MI/NBL since
there is no strict distinction between MI and NBL phases for $T>0$.
Note that, in such a $\rho-T$ phase diagram, the MI/NBL phases get
pinched into exponentially small regions [e.g., in the vicinity of
$\rho=1$ in Fig. \ref{fig:phase12} (d)] as $T\to 0$ and two
zero-temperature tricrtical points get mapped onto each other [e.g.,
TCP01 and TCP02 in Fig. \ref{fig:phase12} (d)].

We now investigate the case $U_2\neq 0$, so the onsite interaction
between bosons depends on the spin. This lifts some of the
infinite degeneracy we encountered in the case $U_2 = 0$ as
can be seen directly at $T=0$ by comparing the pseudo-grayscale
plots of ${\cal E}_0$ in Figs. \ref{fig:lobes} (b), (c), and (d),
for $U_2 = 0$, $U_2 > 0$, and $U_2 < 0$, respectively.

\begin{figure}[htbp]
  \centering
  \epsfig{file=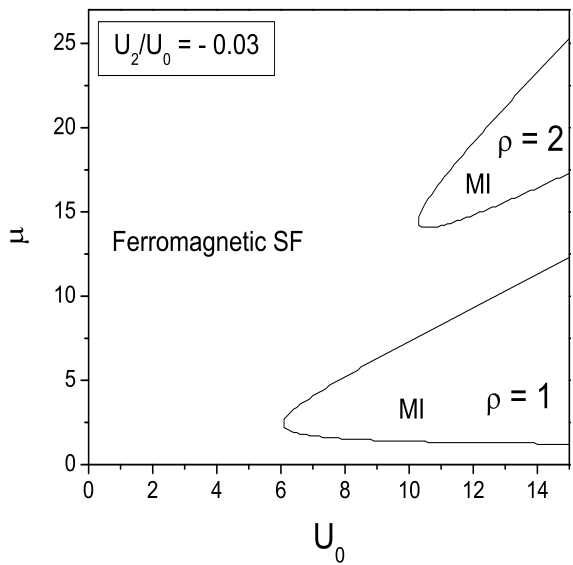,width=9cm,height=9cm}
\caption{Mean-field phase diagram in the $(\mu-U_0)$ plane for
$U_2/U_0=-0.03$ and $T=0$. This has the same topology as the phase
diagram for the case $U_2 = 0$ [Fig. \ref{fig:lobes} (a) for $T=0$]
but the MI lobes have shrunk marginally; the SF-MI transitions are
continuous.} \label{fig:lobes_min}
\end{figure}

\begin{figure}[htbp]
  \centering
  \epsfig{file=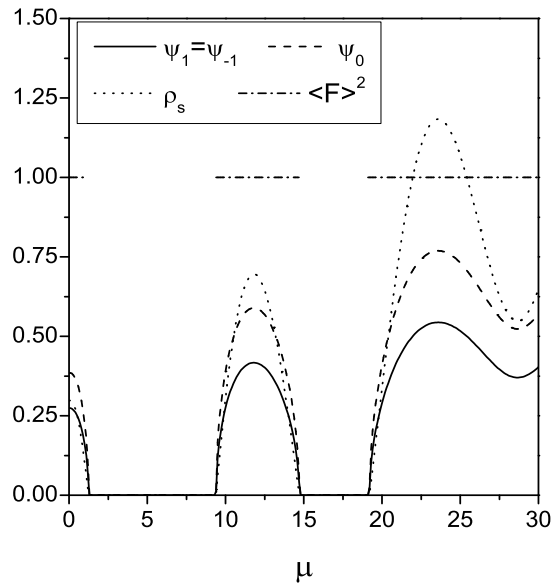,width=9cm,height=9cm}
\caption{Mean-field values of the superfluid order parameters
$\psi_1$=$\psi_{-1}$ and $\psi_0$ and $\rho_s$ plotted as functions
of $\mu$ for $U_0=12$, $U_2/U_0=-0.03$, and $T=0$; $\rho_s$ goes to
zero continuously at the SF-MI transitions. The parameter
$\langle {\vec{F}}\rangle ^2$, defined only in the SF phase,
assumes the value $1$, which confirms that we have a ferromagnetic
SF phase in this case.}
\label{fig:r12_sep_min}
\end{figure}

If $U_2 < 0$ there are four degenerate minima of,
each corresponding
to a ferromagnetic SF, with $\psi_{-1} = \psi_1 $ and
$\psi_0 = \pm\sqrt{2} \psi_{\pm 1}$. The zero-temperature,
mean-field, phase diagram for this case is shown in Fig.
\ref{fig:lobes_min}; it has the same topology as the phase diagram
for the case $U_2 = 0$ [Fig. \ref{fig:lobes} (a)]. We see that
the MI phases have shrunk marginally and the SF-MI transitions are
still continuous. The continuous nature of the $T=0$,
SF-MI transition is illustrated
by the continuous variation of $\psi_{\pm 1}, \psi_0,$
and $\rho_s$ as functions of $\mu$ in Fig. \ref{fig:r12_sep_min}.
The parameter $\langle {\vec{F}}\rangle ^2$, defined only in
the SF phase, assumes the value $1$, which confirms that
we have a ferromagnetic SF phase in this case. The $\mu-T$ phase
diagram for the case $U_2 < 0$ has the same topology as
the $U_2 = 0$ phase diagrams of Fig. \ref{fig:phase12}.
We do not show the $\mu-T$ phase diagram for $U_2 < 0$ since,
for the parameters we use, namely, $U_0 = 12$ and $U_2/U_0 =-0.03$,
the phase boundaries are very close to those in
Fig. \ref{fig:phase12}.

\begin{figure}[htbp]
  \centering
  \epsfig{file=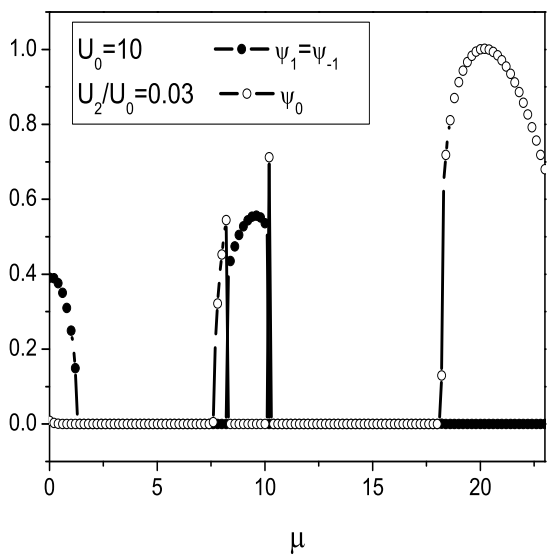,width=9cm,height=9cm}
\caption{Mean-field values of the superfluid order parameters
$\psi_1=\psi_{-1}$ and $\psi_0$ plotted as functions of $\mu$
for $U_0=10$, $U_2/U_0=0.03$, and $T=0$. The SF phase has either
$\psi_{-1}=\psi_1\neq 0$ and $\psi_0=0$ or vice versa, which
confirms that we have a polar SF in this case. }
\label{fig:rho_rhos_10_sep_plus}
\end{figure}

\begin{figure}[htbp]
  \centering
  \epsfig{file=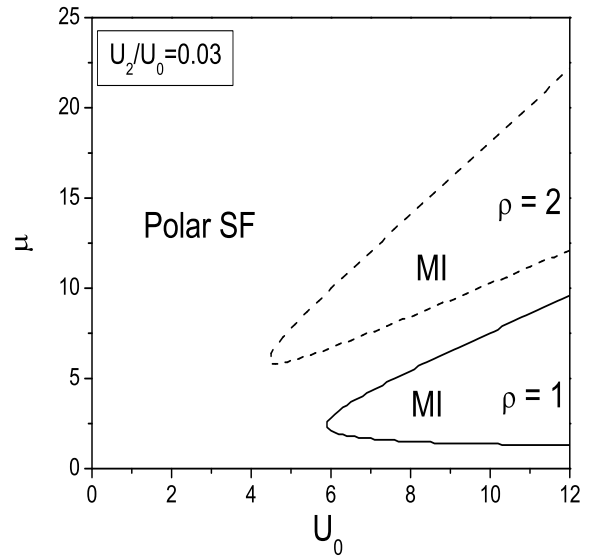,width=9cm,height=9cm}
\caption{Mean-field phase diagram in the $(\mu-U_0)$ plane for
$U_2/U_0=0.03$ and $T=0$. The $\rho=1$ MI lobe has the same form as
its counterpart for $U_2 = 0$ [Fig. \ref{fig:lobes} (a)]. We see that the MI
lobes expand marginally, and the SF-MI transitions are continuous
(represented by a continuous line); but for $\rho=2$ the SF-MI
transition becomes first-order (represented by a dashed line) and
the MI phase is stable over a much wider region of parameter space
than for the case $U_2 =0$.}
\label{fig:lobes_plus}
\end{figure}

\begin{figure}[htbp]
  \centering
  \epsfig{file=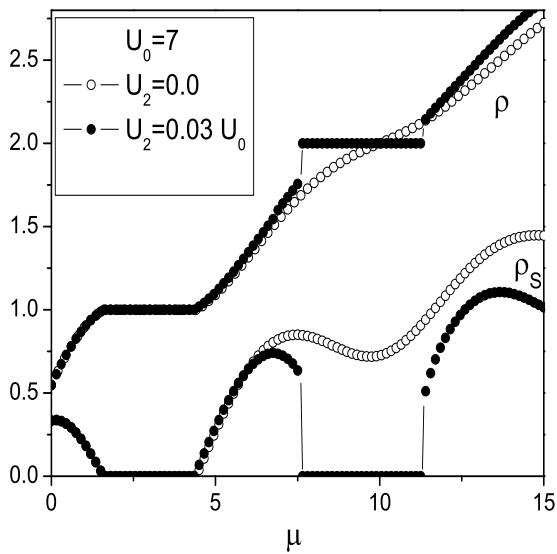,width=9cm,height=9cm}
\caption{Plots of $\rho_s$ and $\rho$ as functions of $\mu$ for
$T=0$, $U_0=7$, and $U_2/U_0=0.03$ (filled circles) and $U_2 = 0$
(open circles). In the former case $\rho_s$ changes continuously at
the SF-MI transitions at the boundary of the $\rho = 1$ MI lobe
(Fig. \ref{fig:lobes_plus}) but jumps at the first-order SF-MI
transitions associated with the boundary of the $\rho=2$ MI lobe.
For $U_2=0$ only the $\rho = 1$ lobe is encountered in this plot and
the SF-MI transitions are continuous; $\rho_s$ shows a gentle
minimum in the vicinity of the $\rho = 2$ MI lobe.}
\label{fig:rhos1}
\end{figure}

If $U_2 > 0$ there are four degenerate minima of the variational
free energy ${\cal F}$ shown, e.g., at $T=0$ in the pseudo-grayscale
plot of Fig. \ref{fig:lobes}. Each one of these minima corresponds
to a polar SF, with either $\psi_{-1}=\psi_1\neq 0$ and $\psi_0=0$
or vice versa as shown in the plots of $\psi_{\pm 1}$ and $\psi_0,$
versus $\mu$ in Fig. \ref{fig:rho_rhos_10_sep_plus} for $U_2/U_0 =
0.03$. The parameter $\langle {F}\rangle $, defined only in the SF
phase, assumes the value $0$, which also confirms that we have a
polar SF phase. The zero-temperature, mean-field, phase diagram for
this case is shown in Fig. \ref{fig:lobes_plus}. If the density
$\rho$ is equal to an odd integer ($\rho=1$ is shown in Fig.
\ref{fig:lobes_plus}), this phase diagram has the same form as its
counterpart for $U_2 = 0$ [Fig. \ref{fig:lobes} (a)]. We see that
the MI lobes expand marginally, and the SF-MI transitions are still
continuous. However, if the density $\rho$ is equal to an even
number ($\rho=2$ is shown in Fig. \ref{fig:lobes_plus}), the SF-MI
transition becomes first-order and the MI phase is stable over a
much wider region of parameter space than for the case $U_2 =0$: As
we show in Fig. \ref{fig:rhos1}, for $U_2/U_0 = 0.03, U_0 = 7$ and
$T=0$, $\rho_s$ and $\rho$ vary continuously as functions of $\mu$
at the SF-MI transition for $\rho=1$ but discontinuously for
$\rho=2$; for comparison we also include the analogous plots for
$U_2 = 0$. [We use $U_0 = 7$ here, rather than $U_0 = 12$, to
compress the range of $\mu$ over which the SF-MI transitions occur.]

\begin{figure*}[htbp]
  \centering
  \epsfig{file=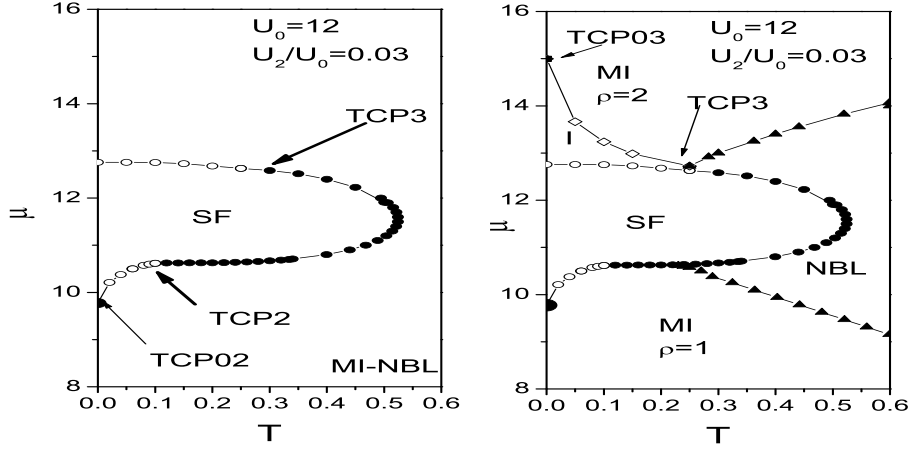,width=14cm,height=7cm}
\caption{(a) Mean-field phase diagram in the $(\mu-T)$ plane for
$U_0=12$, $U_2/U_0=0.03$ showing first-order (open circles)
and continuous (filled circles) transitions between SF and MI/NBL
phases and tricritical points (TCP). This phase diagram is nearly
the same as the $U_2 = 0$ phase diagram [Fig. \ref{fig:phase12}],
but there are no zero-temperature, tricritical points for the
first-order boundaries associated with the MI lobes for even values
of $\rho$: e.g., TCP03 in Fig.  \ref{fig:phase12} has no counterpart
here. (b) A quantitative comparison between these two phase diagrams
shows that the two phase diagrams are nearly indistinguishable
except for the first-order boundaries that link the
zero-temperature, SF-MI transitions for even values of $\rho$ with
the tricritical points directly above them (TCP3 here);
lines with open circles (open diamonds) denote the
first-order boundaries for $U_2/U_0 =0.03$ ($U_2 = 0$); the region
\textbf{I} between these lines lies in the MI (SF) phase if
$U_2/U_0 =0.03$ ($U_2 = 0$); the lines with filled triangles
show the MI-NBL crossover boundary.}
\label{fig:phase3}
\end{figure*}

\begin{figure}[htbp]
  \centering
  \epsfig{file=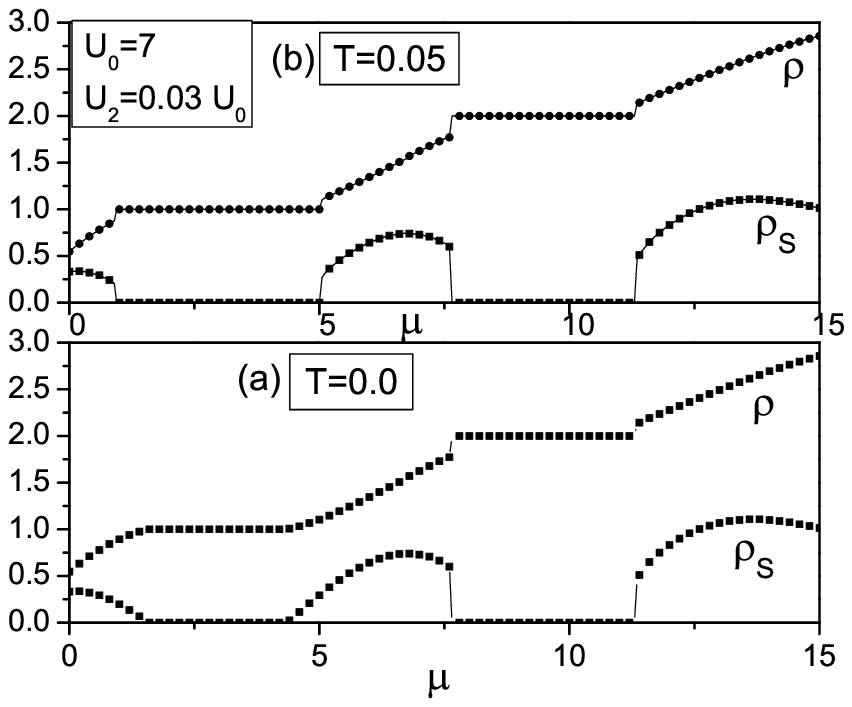,width=8cm,height=8cm}
\caption{Representative plots of $\rho_s$ and $\rho$ versus $\mu$
for (a) $T=0$ and (b) $T=0.05$ for $U_0=7$, $U_2/U_0=0.03$ showing
jumps at first-order SF-MI transitions; $\rho_s$ changes continuously
at the $T=0$, continuous SF-MI transition associated with the
$\rho = 1$ MI lobe.}
\label{fig:rhos1a}
\end{figure}

For $\rho=2$ in the MI phase, there are exactly two bosons localized
per site and the total spin at every site can be either $S=0$ or
$S=2$. Since $U_2 > 0$ there is an energy difference between the
$S=0$ and $S=2$ states, with a lower energy for the singlet state.
To go from the MI to the SF phase, this singlet state has to be
broken by supplying an energy $\sim U_2$, which gives a rough
estimate for the \textit{latent heat} of this first-order
transition if $0 \lesssim T$. This requirement of a latent heat
leads to the greater stability of the MI phases for even values of
$\rho$ relative to their counterparts for odd values of $\rho$.
Thus the $\rho=2$ MI lobe in Fig. \ref{fig:lobes_plus} is
substantially larger than the one for $\rho=1$.
The $\mu-T$ phase diagram for the case $U_2 > 0$, shown in Fig.
\ref{fig:phase3} (a), for $U_0 = 12$ and $U_2/U_0 = 0.03$,
has nearly the same form as the $U_2 = 0$ phase diagram of
Fig. \ref{fig:phase12}. The principal qualitative
difference between these phase diagrams is that, if  $U_2 > 0$,
there are no zero-temperature, tricritical points for the
first-order boundaries associated with the MI lobes for even values
of $\rho$; e.g., the tricritical point TCP03 in Fig.
\ref{fig:phase12} has no counterpart in Fig. \ref{fig:phase3} (a).
A quantitative comparison between these two phase diagrams
is made in Fig. \ref{fig:phase3} (b); this shows that the two
phase diagrams are nearly indistinguishable except for the
first-order boundaries that link the zero-temperature, SF-MI
transitions for even values of $\rho$ with the tricritical
points directly above them (e.g., TCP3). In Fig. \ref{fig:phase3}
(b) the dashed line with open circles (open diamonds) is the
first-order boundary for $U_2/U_0 =0.03$ ($U_2 = 0$); the region
\textbf{I} between these lines lies in the MI (SF) phase if
$U_2/U_0 =0.03$ ($U_2 = 0$); the lines with filled triangles
show the MI-NBL crossover as in Fig. \ref{fig:phase12} (a).
Phase diagrams such as Fig. \ref{fig:phase3} are obtained by
calculating the order parameters $\psi_\sigma$, and thence $\rho$ and
$\rho_s$, as functions of $\mu$ at different temperatures.
Representative plots are shown in Fig. \ref{fig:rhos1a} for
$U_2/U_0 = 0.3, U_0 = 7,$ and $T=0$ and $T=0.05$.

\section{Conclusions}

We have carried out the most extensive study of the phase diagram of
the spin-1 Bose Hubbard model so far by generalizing an intuitively
appealing mean-field theory that has been used earlier for the
spinless case. Our study yields both zero-temperature and
finite-temperature phase diagrams for this model. Only $T=0$ phase
diagrams had been obtained so far \cite{tsuchiya,graham,rvpiacs}; so
our elucidation of the finite-temperature properties of this model
yields qualitatively new insights. We find, in particular, that
several of the SF-MI transitions in this model are generically first
order; at sufficiently high temperatures they become continuous via
tricritical points. Tricritical points also abound at zero
temperature since some, but not all, of the finite-temperature,
first-order transitions become continuous as $T\to0$. The resulting
phase diagrams (Figs. \ref{fig:phase12} and \ref{fig:phase3}) are
very rich and should provide a challenge for experimental studies,
which we hope our work will stimulate. Experiments can study both
the case $U_2 < 0$, which can be realized possibly by using
$^{87}Rb$, and the case $U_2 > 0$, which can be realized by using
$^{23}Na$. Thus, in principle, both the phase diagrams of Figs.
\ref{fig:phase12} and \ref{fig:phase3} could be obtained
experimentally. Of course this will require good experimental
control of both the temperature and the density (or chemical
potential) of the bosons.

Our mean-field theory has been designed to investigate the relative
stabilities of SF and MI/NBL phases. It has enough structure to
unravel the differences between polar and ferromagnetic supefluids.
However, our mean-field theory does not account for order parameters
that can distinguish between different spin orderings in the MI
phase, e.g., spin-singlet and nematic MIs, which have been
investigated in the limit $U_0\to\infty$ by some
groups\cite{imambekov,fath}. The generalization of our mean-field
theory to include such types of structures in the MI phases of model
(\ref{eq:bh}) lies beyond the scope of this study but is an
interesting challenge for further theoretical work.

\begin{acknowledgments}
One of us (RVP) thanks the Jawaharlal Nehru Centre for Advanced
Scientific Research and the Department of Physics, Indian Institute
of Science, Bangalore for hospitality during the time when a part of
this paper was written. This work was supported by DST, India
(Grants No. SP/S2/M-60/98 and SP/I2/PF-01/2000) and UGC, India.
\end{acknowledgments}


\begin{thebibliography}{11}
\bibitem{foot1} We have checked that this value of $n_{max}$
suffices for the large values of $U_0$ we use by comparing our
results with those for $n_{max} = 2$ and $3$.
\end{thebibliography}
\end{document}